\documentclass[a4paper,12pt]{article}
\usepackage[francais]{babel}
\usepackage[latin1]{inputenc}
\usepackage[OT1]{fontenc}
\usepackage{graphicx}
\usepackage{lfa06}  
\usepackage{times}
\usepackage{subfigure,amssymb}

\newcommand{\ocap}{\textcircled{$\scriptstyle{\cap}$}}
\newcommand{\ocup}{\textcircled{$\scriptstyle{\cup}$}}

\begin{document}

\renewcommand{\figurename}{Figure~}
\renewcommand{\tablename}{Tableau~}
\date{}

\title{Un filtre temporel crédibiliste pour la reconnaissance d'actions humaines dans les vidéos \\[1.3em]
A temporal belief filter for human action recognition in videos }

\author{\small
        \begin{tabular}[t]{c}
        E. Ramasso, M. Rombaut, D. Pellerin
        \end{tabular}
        ~\\
\small{Laboratoire des Images et des Signaux (LIS - UMR 5083)}
        ~\\
        \\
        \small{46, avenue Félix Viallet, 38031 Grenoble, France,}\\
\small{\{emmanuel.ramasso,michele.rombaut,denis.pellerin\}@lis.inpg.fr}
}

\parskip 3mm
\maketitle

\thispagestyle{empty}

\doubleresume{Ce papier présente un filtre temporel crédibiliste
utilisé pour la reconnaissance d'actions humaines dans des vidéos.
Ce filtre permet de s'affranchir au moins en partie des problèmes
dus à la disparité dans la réalisation des actions, à la
variabilité des conditions d'acquisition des vidéos et à la
difficulté d'appliquer les algorithmes de traitements d'images sur
des vidéos réelles. L'ensemble du système de reconnaissance est
construit à partir du formalisme du Modèle des Croyances
Transférables (Transferable Belief Model : TBM) proposé par P.
Smets. Le TBM permet d'exprimer explicitement le doute entre les
actions. De plus, l'information de conflit mise en lumière par le
TBM est exploitée pour détecter le changement d'état des actions.
Les performances du filtre sont estimées sur des vidéos
d'athlétisme réelles acquises en caméra mobile avec  des angles de
vue variables.} {Filtrage de fonctions de croyance, Modèle des
Croyances Transférables, Indexation de vidéos.} {In the context of
human action recognition in video sequences, a temporal belief
filter is presented. It allows to cope with human action disparity
and low quality videos. The whole system of action recognition is
based on the Transferable Belief Model (TBM) proposed by P. Smets.
The TBM allows to explicitly model the doubt between actions.
Furthermore, the TBM emphasizes the conflict which is exploited
for action recognition. The filtering performance is assessed on
real video sequences acquired by a moving camera and under several
unknown view angles.} {Belief functions filtering, Transferable
Belief Model, video indexing.}

\section{Introduction}
\subsection{Contexte}
L'analyse du comportement humain dans les vidéos est un domaine de
recherche en plein essor dans la communauté de la vision par
ordinateur~\cite{Wang03a}. En effet, elle est au coeur de nombreuses
applications telles que la surveillance de personnes, les
interfaces homme-machine, l'indexation et la recherche dans
de grandes bases de vidéos. Le lien entre le monde réel, de
nature analogique, et la pensée humaine, plutôt symbolique, est une des
difficultés majeures.
%

De nombreuses méthodes sont proposées pour la reconnaissance
d'actions humaines dans les vidéos~\cite{Wang03a}. Elles peuvent
être classées en deux catégories : les \emph{templates matching}
et les \emph{machines à états}. Le premier  type de méthodes ne
parvient généralement pas à faire face aux conditions variables
d'acquisition des vidéos telles que le changement d'angle de vue
et la disparité dans la réalisation des actions. La deuxième
catégorie concerne majoritairement des méthodes
probabilistes~\cite{Hongeng04} notamment les \emph{Modèles de
Markov Cachés} (HMM) et les \emph{Réseaux Bayésiens Dynamiques}
(DBN)~\cite{Luo03}. La théorie des probabilités est
particulièrement intéressante lorsque des jeux de données
d'apprentissage conséquents~\cite{Freitas03} sont disponibles et
lors de la prise de décision.

L'analyse du mouvement humain basée sur les fonctions de croyance est une approche récente.
Un classifieur basé sur le modèle de Shafer~\cite{shafer76a} a été utilisé pour reconnaître des
postures statiques~\cite{Girondel05} et des expressions faciales~\cite{Hammal05}. Nous avons
proposé~\cite{Ramasso05} une architecture originale pour la reconnaissance d'actions humaines basée sur le
Modèle de Croyances Transférables (Transferable Belief Model : TBM)~\cite{Smets94}.
Ce formalisme est particulièrement bien adapté car
(i) le \emph{doute} sur les états est modélisé explicitement, (ii) le \emph{conflit} entre les sources
d'information peut être utilisé pour détecter un changement d'état et pour remettre en question les sources
fusionnées et (iii) la \emph{fiabilité} des paramètres dépendant du contexte peut être prise en compte.

Les méthodes proposées jusqu'ici pour l'analyse du mouvement humain basée sur les fonctions de croyance
n'intégrent pas l'information temporelle. Pour remédier à cela, deux solutions principales existent :
les réseaux évidentiels~\cite{Xu94} et les réseaux de Petri crédibilistes~\cite{Mrecsqaru99}.
Cependant, ces deux méthodes sont sensibles au bruit pouvant apparaître sur les croyances et se
traduisant généralement par la présence de fausses alarmes.
De plus, les fonctions de croyance utilisées sont généralement normalisées pour supprimer le conflit.

Dans cet article, nous proposons une méthode de filtrage des
fonctions de croyance qui est capable d'éliminer les fausses
alarmes et qui exploite l'information de conflit.

\subsection{Reconnaissance des actions}

Indexer une vidéo à un niveau compatible avec la compréhension humaine nécessite de
définir des concepts c'est à dire des informations dites de haut niveau symbolique. Nous nous
intéressons ici au comportement d'un athlète au cours de meetings d'athlétisme
et nous cherchons à détecter et reconnaître ses actions comme la \emph{course}, le \emph{saut} et
la \emph{chute}. Ces actions sont supposées indépendantes, non exhaustives et
non exclusives.

Une action peut prendre deux états distincts : soit elle est vraie
soit elle fausse. Pour déterminer l'état d'une action, des
paramètres sont extraits du flux vidéo par des méthodes de
traitements classiques comme l'estimation du mouvement de caméra
et le suivi de points. Les paramètres sont choisis pour leur
pertinence par rapport aux actions à reconnaître. Dans le cadre de
l'application avec caméra mobile, notre choix s'est basé sur deux
hypothèses principales : (i) le cameraman suit l'athlète et (ii)
la trajectoire de points particuliers de la silhouette suffit à
décrire le comportement de l'athlète. Les paramètres qui ont été
choisis sont : les mouvements de caméra entre deux images
successives (horizontal, vertical, zoom) et la trajectoire de la
tête, d'un des deux pieds et du centre de la silhouette. Ces
derniers paramètres permettent d'extraire l'alternance des pieds,
la courbure de la silhouette et son angle par rapport à l'horizon.
Tous les paramètres sont ensuite traduits en croyance puis
fusionnés dans le cadre du TBM. Finalement, à chaque image, une
croyance sur la réalisation de chaque action est calculée.

\begin{figure}[!ht]
\center
\includegraphics[width=7cm]{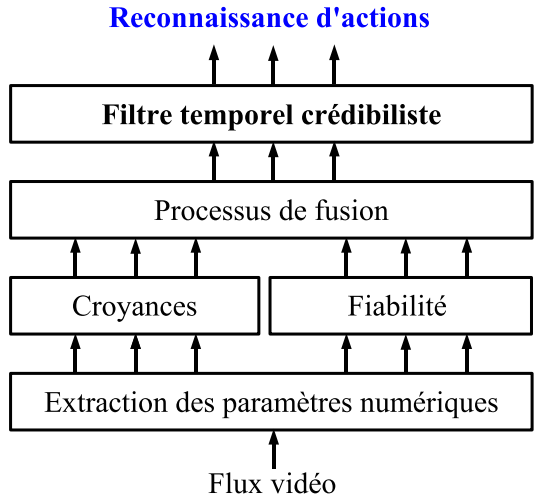}
\caption{Architecture du système de reconnaissance d'actions.
\label{fig:archi}}
\vspace{-0.1cm}
\end{figure}

Du fait de la diversité dans la réalisation des actions, des
conditions d'acquisition des vidéos et de la complexité des
algorithmes de traitement, les résultats de la fusion sont
bruités. Il est bien sûr possible de filtrer les signaux
numériques, mais cela pourrait éliminer des informations utiles
aux actions notamment lorsque celles ci ne durent que quelques
images. Nous avons donc développé un filtre temporel crédibiliste
capable de filtrer le bruit sur les croyances mais aussi de
détecter les changements d'états pour chacune des actions. Le
système complet est présenté figure \ref{fig:archi}.

Dans le paragraphe \ref{sec:fusion}, nous décrivons comment
l'information concernant une action particulière est modélisée
puis fusionnée dans le cadre du TBM. Le paragraphe \ref{sect:filt}
est consacré à la description du filtre temporel crédibiliste 
que nous avons développé. Enfin, le filtre
est évalué dans le paragraphe \ref{sec:result}.

\section{Processus de fusion}
\label{sec:fusion}

\subsection{Définition des fonctions de croyance}

Soit $\Omega_A=\{R_A,F_A\}$ l'espace de discernement (FoD) d'une action $A$.
Le FoD regroupe les états élémentaires d'une action (vrai ou faux).
Une distribution de masse $m^{\Omega_A}_{P}$ concernant une action $A$,
est une fonction de croyance dépendant du paramètre $P$ définie sur
l'ensemble des propositions $2^{\Omega_A}=\{\emptyset, R_A,F_A,R_A
\cup F_A\}$ (où $R_A \cup F_A$ correspond au doute entre les états
$R_A$ et $F_A$). La distribution de masse est une application qui
à tout $X \in 2^{{\Omega}_A}$ associe $m^{{\Omega}_A}_{P}(X) \in [0,1]$. Par
construction $m^{\Omega_A}_{P}(\emptyset)=0$, et $\sum_{X \subseteq \Omega_A}
m^{\Omega_A}_{P}(X)=1$. La valeur de
$m^{\Omega_A}_{P}(X)$ correspond à la masse qui indique la
confiance sur la proposition $X \subseteq \Omega_A$ à partir de la
connaissance du paramètre $P$. Cette valeur ne donne aucune information
supplémentaire concernant les sous ensembles de $X$. C'est la
différence fondamentale avec une mesure de probabilité.

Nous nous sommes inspirés des sous ensembles flous~\cite{Ramasso05} pour obtenir
les distributions de masse de chaque paramètre numérique. Cette conversion
numérique-symbolique est réalisée à chaque image de la vidéo analysée.

\subsection{Fusion des fonctions de croyance}

Les distributions de masses obtenues à partir des différents paramètres sont
fusionnées dans le cadre du TBM~\cite{Smets94}.
Une fonction de croyance est alors obtenue pour chacune des actions possibles prenant en compte
toute l'information disponible. La fusion est réalisée image par image et pour chaque action.

La combinaison de deux distributions de masse $m_{P_1}^{\Omega_A}$ et $m_{P_2}^{\Omega_A}$
définies sur le même FoD $\Omega_A$, concernant une même action $A$
et calculées à partir de la valeur numérique de deux paramètres $P_1$ et $P_2$, donne :
\begin{eqnarray}
m_{P_1}^{\Omega_A} {\textcircled{$\triangle$}}
m_{P_2}^{\Omega_A}(E)  =  \sum_{C \triangle D = E}
m_{P_1}^{\Omega_A}(C).m_{P_2}^{\Omega_A}(D)
\label{eq:comb}
\end{eqnarray} avec $\triangle=\cap$ (resp. $\cup$) pour
la combinaison conjonctive (resp. disjonctive). Une action est
alors décrite par des règles logiques entre les états des
paramètres (exemple : ``\emph{si} l'alternance des pieds est
importante \textsc{et} si la translation horizontale de la caméra
est importante \emph{alors} l'action est une course") puis
retranscrites dans le cadre du TBM où le \textsc{et} logique est
remplacé par $\ocap$ et le \textsc{ou} logique par $\ocup$ en
prenant soin d'avoir exprimé les fonctions de masse sur le même
FoD.

\subsection{Fiabilité des sources}

Il est possible de prendre en compte la fiabilité des
paramètres lors de la combinaison. Cela permet de
modérer l'influence d'une source d'information pour laquelle les conditions de
fonctionnement ne seraient pas optimales. Le paramètre de fiabilité $\alpha_P \in
[0,1]$ concernant le paramètre $P$ agit sur la distribution de masse $m_P^{\Omega_P}$ de la façon suivante :
\begin{equation}\begin{array}{lll}
m_P^{\Omega_P,\alpha_P}(A) & =  \alpha_P \times m_P^{\Omega_P};
\forall A \subsetneq {\Omega_P}\\
m_P^{\Omega_P,\alpha_P}(\Omega_P) & =  (1-\alpha_P) + \alpha_P
\times m_P^{\Omega_P} &\end{array}\nonumber\end{equation}
L'estimation de $\alpha_P$ peut être réalisée à partir de connaissances statistiques~\cite{Elouedi}.
Nous proposons ici de calculer en ligne le facteur de fiabilité lié à
la qualité de la vidéo et des traitements associés à chaque image.
Deux facteurs de fiabilité ont été définis. Tout d'abord
$\alpha_{sup}$ qui est basé sur l'appartenance effective du
pixel au mouvement dominant et qui permet de quantifier la qualité de
l'estimation de mouvement. Ensuite, $\alpha_{dist}$ qui
est calculé à partir du rapport normalisé entre
les distances \emph{pieds - centre de gravité} et \emph{tête -
centre de gravité} dont la valeur optimale est 1. Ce deuxième facteur
permet d'évaluer la qualité de l'algorithme de suivi de points de la silhouette.

\section{Filtre temporel crédibiliste}\label{sect:filt}

Les fonctions de croyance obtenues par fusion des paramètres sont
traitées par un filtre temporel crédibiliste qui permet de lisser
l'évolution des croyances et de détecter les changements d'états.
Ce filtre est appliqué sur les croyances de chacune des actions
indépendamment.

En sortie du filtre, une distribution de masse est obtenue. Elle est cohérente avec :\\
\begin{itemize}
\item \emph{les paramètres} : le
filtre permet de résoudre le conflit entre les sources associées à
chacun des paramètres,\\
\item \emph{les variations dans le temps} : la croyance sur les actions ne peut varier
brutalement d'une image à l'autre compte tenu de la différence
entre la cadence vidéo et les mouvements humains,\\
\item \emph{la condition d'exclusivité} : le
filtre permet de s'assurer de la consonance des distributions de
masse c'est à dire que si l'une des hypothèses, e.g. $R_A$, a une masse non nulle
alors l'autre hypothèse, e.g. $F_A$, a une masse nulle.\\
\end{itemize}

\begin{figure}[ht]\center
\includegraphics[width=\columnwidth]{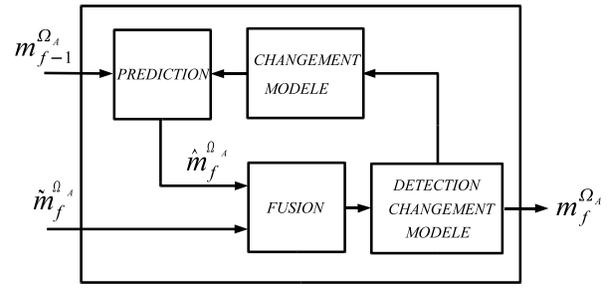}
\caption{Principe du filtre temporel crédibiliste où
$\hat{m}_{f}^{\Omega_A}$ est la prédiction, $m^{\Omega_{A}}_f$ est
la valeur de sortie du filtre à l'image $f$ donnant l'état de
l'action et $\tilde{m}_{f}^{\Omega_A}$ est la mesure en sortie de
la fusion de paramètres à l'image $f$.\label{fig:belfiltpr}}
\end{figure}

La dernière propriété permet de définir un "état" de l'action, i.e.
soit \emph{vrai} soit \emph{faux}. Ces états sont évalués pour
chaque action \emph{indépendamment} en fonction des distributions de masse
définies à partir de chaque paramètre.

Le principe du filtre temporel crédibiliste est décrit
figure~\ref{fig:belfiltpr}. Le filtre temporel est basé sur les
\emph{règles d'implication}. La formalisation de ces règles dans
le cadre du TBM ainsi qu'un exemple concernant l'identification de
cibles aériennes sont fournis dans~\cite{Ristic04}. Nous utilisons
deux règles d'implication que nous interprétons comme des modèles
d'évolution des masses sur les actions. Nous notons les modèles
$\mathcal{M} \in \{\mathcal{R},\mathcal{F}\}$ avec $\mathcal{R}$
pour le modèle \emph{vrai} et $\mathcal{F}$ pour le modèle
\emph{faux}. Chacun de ces modèles concerne l'une des hypothèses
du FoD d'une action $A$, i.e. $R_A$ ou $F_A$. A chaque image $f$
($f$ pour frame), le filtre fonctionne en trois étapes : (i)
prédiction, (ii) fusion et (iii) détection de changement d'état.

\subsection{Prédiction}

L'étape de prédiction s'appuie sur l'hypothèse suivante : si l'état de l'action est $R_A$ (resp. $F_A$) à l'image
$(f-1)$ alors, à l'image $f$ son état sera au moins partiellement $R_A$ (resp. $F_A$). Ce modèle d'évolution
$\mathcal{R}$ (resp. $\mathcal{F}$) est pondéré par un coefficient de confiance $\gamma_\mathcal{R} \in [0,1]$
(resp. $\gamma_\mathcal{F} \in [0,1]$) :
\begin{equation}\begin{array}{l}
\textbf{\textrm{Modèle $\mathcal{R}$ :}} \\
\hspace{0.8cm}\textrm{Si} \; R_A \; \textrm{à } (f-1)\\
\hspace{0.8cm}\textrm{ alors} \; R_A \;
\textrm{à $f$ avec la croyance} \;\gamma_{R}\\
\textbf{\textrm{Modèle $\mathcal{F}$ :}}  \\
\hspace{0.8cm}\textrm{Si} \; F_A \;\, \textrm{à } (f-1)\\
\hspace{0.8cm}\textrm{alors} \; F_A \;\, \textrm{à $f$ avec la croyance} \; \gamma_{F}
\end{array}\label{eq:rule}
\end{equation}
Par la suite, la notation vectorielle de la distribution de
masse définie sur le $\Omega_A$ est utilisée comme suit :
$$\begin{array}{l}
m^{\Omega_{A}} = \\
\hspace{-0.1cm}[ m^{\Omega_{A}}(\emptyset)  \; \; \; m^{\Omega_{A}}(R_A)  \; \; \; m^{\Omega_{A}}(F_A)  \; \; \; m^{\Omega_{A}}(\Omega_A)]^T
\end{array}$$

Le modèle d'évolution peut être interprété comme une distribution de masse. Par exemple, pour le modèle
$\mathcal{R}$ :
\begin{equation}m^{\Omega_{A}}_{\mathcal{R}} = \left[
\begin{array}{cccc} 0 & \gamma_{\mathcal{R}} & 0 &
(1-\gamma_{\mathcal{R}}) \end{array} \right]^T\label{eq:rr}
\end{equation}
La règle de combinaison disjonctive
(eq.~\ref{eq:comb}) est alors utilisée pour calculer la prédiction à partir de la distribution de masse à l'image
$(f-1)$ et du modèle d'évolution :
\begin{equation}
\hat{m}_{f,\mathcal{M}}^{\Omega_A} = m^{\Omega_{A}}_{\mathcal{M}}
\ocup {m}^{\Omega_A}_{f-1}\label{eq:predi}
\end{equation}
La règle $\ocup$ n'attribue jamais plus de masse à une hypothèse singleton que celle qu'elle avait avant fusion.
Par conséquence, la distribution de masse prédite à l'image $f$ à l'aide du modèle $\mathcal{R}$ (eq.~\ref{eq:rr})
est donc :
\begin{equation}
\hat{m}_{f,\mathcal{R}}^{\Omega_A}
\hspace{-0.08cm}=\hspace{-0.08cm}
\left[\hspace{-0.28cm}\begin{array}{c}0 \\
\gamma_\mathcal{R} \times m_{f-1}^{\Omega_A}(R_A)\\0 \\
(1-\gamma_\mathcal{R}) \times m_{f-1}^{\Omega_A}(R_A) +
m_{f-1}^{\Omega_A}(\Omega_A)\end{array}\hspace{-0.27cm}\right]
\label{eq:mr}
\end{equation}
Quand $\gamma_\mathcal{M}=1$, la prédiction à $f$  est égale à
la distribution de masse à $(f-1)$. Quand
$\gamma_\mathcal{M}=0$, le modèle ne donne aucune indication sur
l'évolution de la masse entre une image et la suivante.

\subsection{Fusion}

Prédiction $\hat{m}_{f,\mathcal{M}}^{\Omega_A}$ et mesure
$\tilde{m}_{f}^{\Omega_A}$ sont deux sources d'information
concernant l'état de l'action $A$. La combinaison conjonctive des
distributions de masse associées (eq.~\ref{eq:comb}) donne une
nouvelle distribution de masse dont la masse sur l'ensemble vide
quantifie le conflit entre modèle et données et donc reflète le besoin de
changer de modèle :
\begin{eqnarray}\label{eq:err}
    \epsilon _{f}=(\hat{m}_{f,\mathcal{M}}^{\Omega_A} \ocap
\tilde{m}^{\Omega_A}_{f})(\emptyset)
\end{eqnarray}
Cependant, $\epsilon _f$ ne peut pas être utilisé directement car il
peut provenir des erreurs de mesures sur les paramètres.
Nous avons donc utilisé la somme cumulée (\textsc{cusum}) pour
résoudre ce problème et éviter les changements d'état
intempestifs. La \textsc{cusum} est bien adaptée
pour traiter les \emph{changements rapides et importants} ou
\emph{changements longs et graduels}.

\subsection{Détection du changement d'état}

Lorsque la valeur de la \textsc{cusum} atteint un
seuil  d'alerte $\mathcal{T}_w$ (\underline{\textbf{w}}arning threshold),
le numéro de l'image $f_w$ est mémorisé mais le
modèle d'évolution courant reste valide. Quand la \textsc{cusum}
s'accroît et atteint un seuil d'arrêt $\mathcal{T}_s$ (\underline{\textbf{s}}top
threshold) à l'image $f_s$, alors le modèle
d'évolution est changé, puis le nouveau modèle est appliqué à partir de
$f_s$. Lorsque du conflit apparaît, les mesures ne sont pas prises en compte, seule la
valeur prédite est utilisée. Cela revient à \emph{faire confiance au modèle
d'évolution}. Ceci permet de ne pas maintenir une masse sur le
conflit qui est un élément absorbant par la combinaison conjonctive $\ocap$ :
\begin{equation}
{m}^{\Omega_A}_{f} =\left\{\begin{array}{lll}
\hat{m}_{f,\mathcal{M}}^{\Omega_A} \ocap \tilde{m}^{\Omega_A}_{f}
&
&\textrm{si} \;\; \epsilon_f = 0\\& & \\
                    \hat{m}_{f,\mathcal{M}}^{\Omega_A}
&   & \textrm{sinon}
\end{array} \right.
\label{eq:filtass}
\end{equation}
L'équation~(\ref{eq:filtass}) tient compte du fait que la
distribution de masse ${m}^{\Omega_A}_{f-1}$ possède au plus
deux éléments focaux (eq.~\ref{eq:mr}) en fonction du modèle
d'évolution courant $\mathcal{M}$. Au final, la distribution de
masse est consonante, sans conflit et seulement l'une des deux
hypothèses $R_A$ ou $F_A$ a une masse non nulle. De plus, le fait
d'utiliser  la règle disjonctive $\ocup$ pour prédire l'état
permet, en cas de conflit avec les mesures, d'aboutir à l'infini à
une distribution de masse ${m}^{\Omega_A}_{f\to\infty}(\Omega_A)=1$ qui
reflète l'ignorance totale ce qui est tout à fait cohérent.

Tel qu'il a été décrit, le traitement de la \textsc{cusum} pose un
problème pour les conflits très faibles mais de longue durée. Pour
y remédier, nous proposons d'utiliser un mécanisme d'\emph{oubli}
qui permet d'atténuer l'effet du conflit qui serait intervenu
depuis trop longtemps : \textsc{cusum} courante $\mathbf{CS}(f)$ :
\begin{equation}
\mathbf{CS}(f) \leftarrow \mathbf{CS}(f-1) \times \lambda +
\epsilon_f
\end{equation}
Le coefficient d'oubli $\lambda$ a été choisi constant et est appliqué à chaque image.\\

Seul l'un des deux modèles ($\mathcal{R}$ et $\mathcal{F}$) est appliqué à chaque image.
Lorsqu'un changement de modèle intervient pour l'action $A$, c'est à dire que le seuil
d'arrêt $\mathcal{T}_{s}$ est atteint par la \textsc{cusum}, l'intervalle des images $\mathbf{IT}=[f_w,
min(f_s,f_w+\mathcal{W})]$ est interprété comme un intervalle de transition entre les deux états de l'action
$A$. Le paramètre $\mathcal{W}$ limite la taille de la transition. Une distribution de masse modélisant l'ignorance
est affectée en sortie du filtre sur tout l'intervalle $\mathbf{IT}$ :
$m_{\mathbf{IT}}^{\Omega_A}(\Omega_A)=1$.
Puis, à partir de la borne supérieure de $\mathbf{IT}$, la \textsc{cusum} est remise à zéro et le
nouveau modèle, en concordance avec les mesures, est appliqué.

\subsection{Quelques indications sur l'initialisation et le réglage des paramètres}

\subsubsection{Initialisation}

Le filtre temporel crédibiliste est un mécanisme qui fonctionne en
ligne. Lors de l'initialisation, le modèle qui correspond le mieux
aux premières mesures est sélectionné. Pour cela, le calcul de la
\textsc{cusum} est réalisé sur les premières images pour les deux
modèles puis celui présentant la plus petite \textsc{cusum} est
choisi.

\subsubsection{Réglages des paramètres}

Il est nécessaire de configurer les paramètres dans le
bon ordre : le paramètre d'oubli $\lambda$ ainsi que les deux coefficients associés aux modèles $\gamma_\mathcal{R}
\in [0,1]$ (resp. $\gamma_\mathcal{F}$) doivent être traités ensemble et en premier lieu.
Puis le seuil d'arrêt $\mathcal{T}_s$ de la \textsc{cusum} ainsi que le seuil
d'alerte $\mathcal{T}_w$ sont réglés. Enfin le paramètre $\mathcal{W}$ est fixé.

Pour un paramètre d'oubli donné, la valeur de $\mathcal{T}_s$ peut être estimée si une vérité
terrain concernant la validité des actions est disponible c'est à dire si l'image de départ
$f_{sref}$ (\underline{\textbf{s}}tart frame) et l'image d'arrêt $f_{eref}$ (\underline{\textbf{e}}nd frame)
sont connues. Dans ce cas, le filtre est appliqué avec le modèle $\mathcal{F}$ (\emph{état faux}) avec un seuil
d'arrêt $\mathcal{T}_s$ inatteignable. Alors, le seuil vaut ${\mathcal{T}_s}= \mathbf{CS}(f_{sref})$.
S'il y a beaucoup de bruit sur les données, alors il se peut
que la valeur estimée ne soit pas optimale.
Dans ce cas, le paramètre d'oubli $\lambda$ doit être augmenté et
la procédure réitérée.

\section{Expérimentations}
\label{sec:result}

\subsubsection{Description de la base de test}

\begin{figure}[ht]
\begin{center}
\includegraphics[width=\columnwidth]{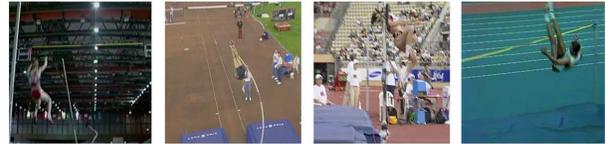}
 \caption{Exemples illustrant l'hétérogénéité de la base de vidéos.}
\label{fig:database}
\end{center}
\vspace{-0.5cm}
\end{figure}

Nous avons utilisé ce système pour reconnaître les actions d'un
athlète dans différents meetings d'athlétisme. Les vidéos
utilisées concernent des \emph{sauts en hauteur} et des
\emph{sauts à la perche}. Les actions recherchées sont la
\emph{course}, le \emph{saut} et la \emph{chute}. La base de
données est composée de 34 vidéos filmées par une caméra mobile.
Il y a $22$ sauts à la perche et $12$ sauts en hauteurs soit
environ $5500$ images. Cette base est caractérisée par son
hétérogénéité~(fig.~\ref{fig:database}) : angles de vue divers,
environnements extérieurs ou intérieurs, présence d'autres
personnes ou d'objets mobiles dans la scène, athlètes hommes ou
femmes de différentes nationalités.

\subsubsection{Réglages}

 Pour montrer les performances et la robustesse du filtre temporel crédibiliste, un seul réglage
a été réalisé pour toutes les actions et toutes les vidéos en utilisant les recommandations
décrites au paragraphe~\ref{sect:filt}. Pour information, les valeurs choisies sont : $\lambda=0.9$,
$\gamma_\mathcal{R}=\gamma_\mathcal{F}=0.9$, $\mathcal{T}_s=3$,
$\mathcal{T}_w=0.5$ et $\mathcal{W}=5$.

\subsubsection{Critère de décision}

 Afin d'évaluer les capacités du
système proposé, il est nécessaire de prendre une décision à partir des croyances sur l'état des actions.
Pour cela, nous utilisons la transformation pignistique proposée par Smets, qui vaut dans notre cas :
\vspace{-0.2cm}
\begin{equation}
\displaystyle
\mathbf{BetP}(R_A)=
\frac{{m_f^{\Omega_A}(R_A)} + 0.5\times m_f^{\Omega_A}(R_A \cup F_A)}{1-m_f^{\Omega_A}(\emptyset)}
\end{equation}
avec $m_f^{\Omega_A}$ la sortie du système avec et sans filtre.
Si $\mathbf{BetP}(R_A)>0$ alors $A$ est considérée comme étant vraie.

\subsubsection{Critère d'évaluation}

La base a été annotée manuellement ce qui nous sert de référence.
Les indices de rappel ($\mathcal{R}$) et précision ($\mathcal{P}$)
sont calculés de la façon
suivante : $\mathcal{R} = \frac{C\cap R}{C}$ et $\mathcal{P} =
\frac{C \cap R}{R}$, où $C$ est l'ensemble des images correctes,
i.e. les références données par l'annotation de l'expert, $R$ est
l'ensemble des images retrouvées à l'aide du critère sur la
probabilité pignistique $\mathbf{BetP}$, et $C \cap R$ est le
nombre d'images correctes retrouvées.

\subsubsection{Illustration}

\begin{figure*}[ht]
\begin{center}
\subfigure{\includegraphics[width=11.5cm]{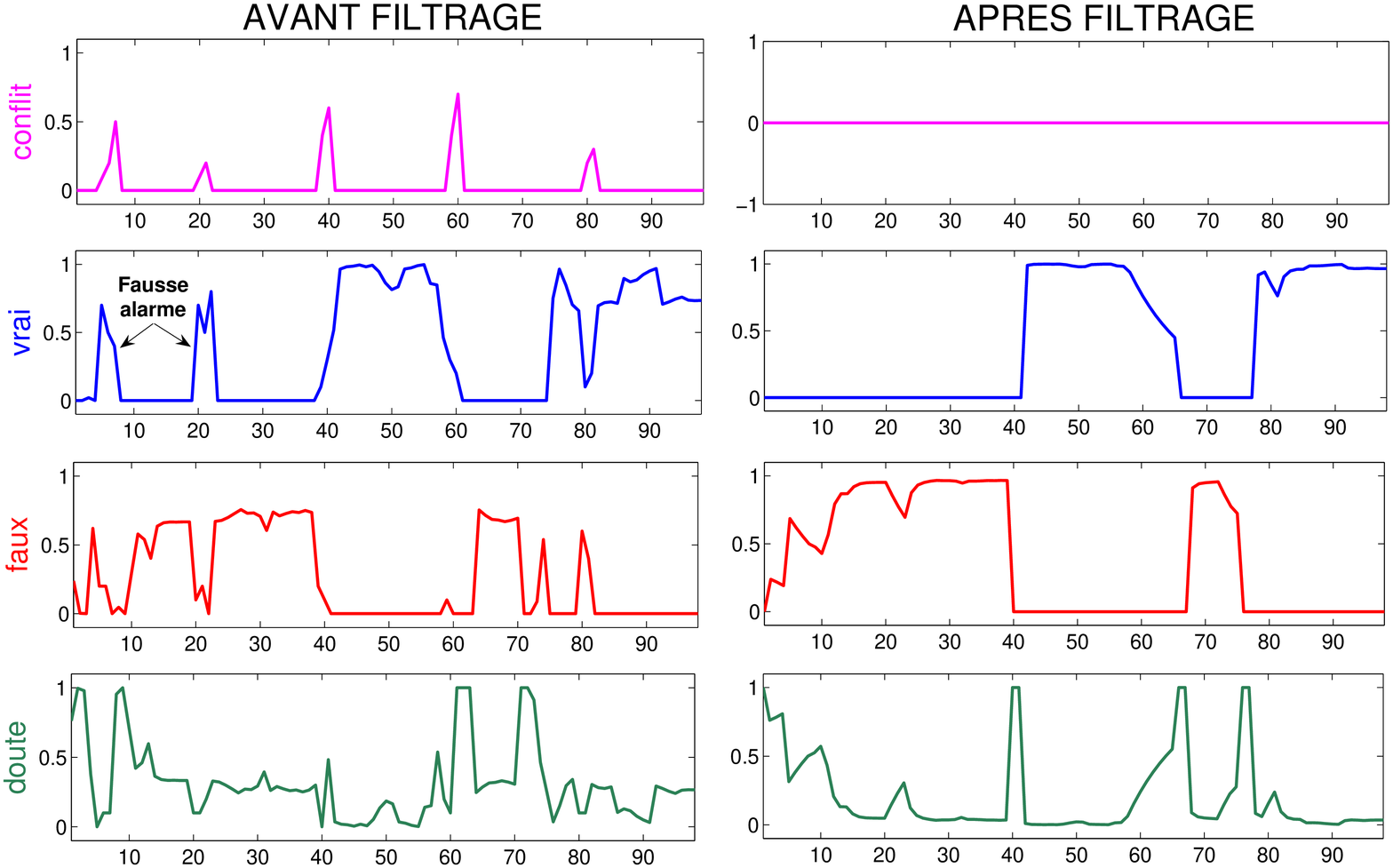}}\vspace{-1cm}
\subfigure{\includegraphics[width=8.5cm,height=3.2cm]{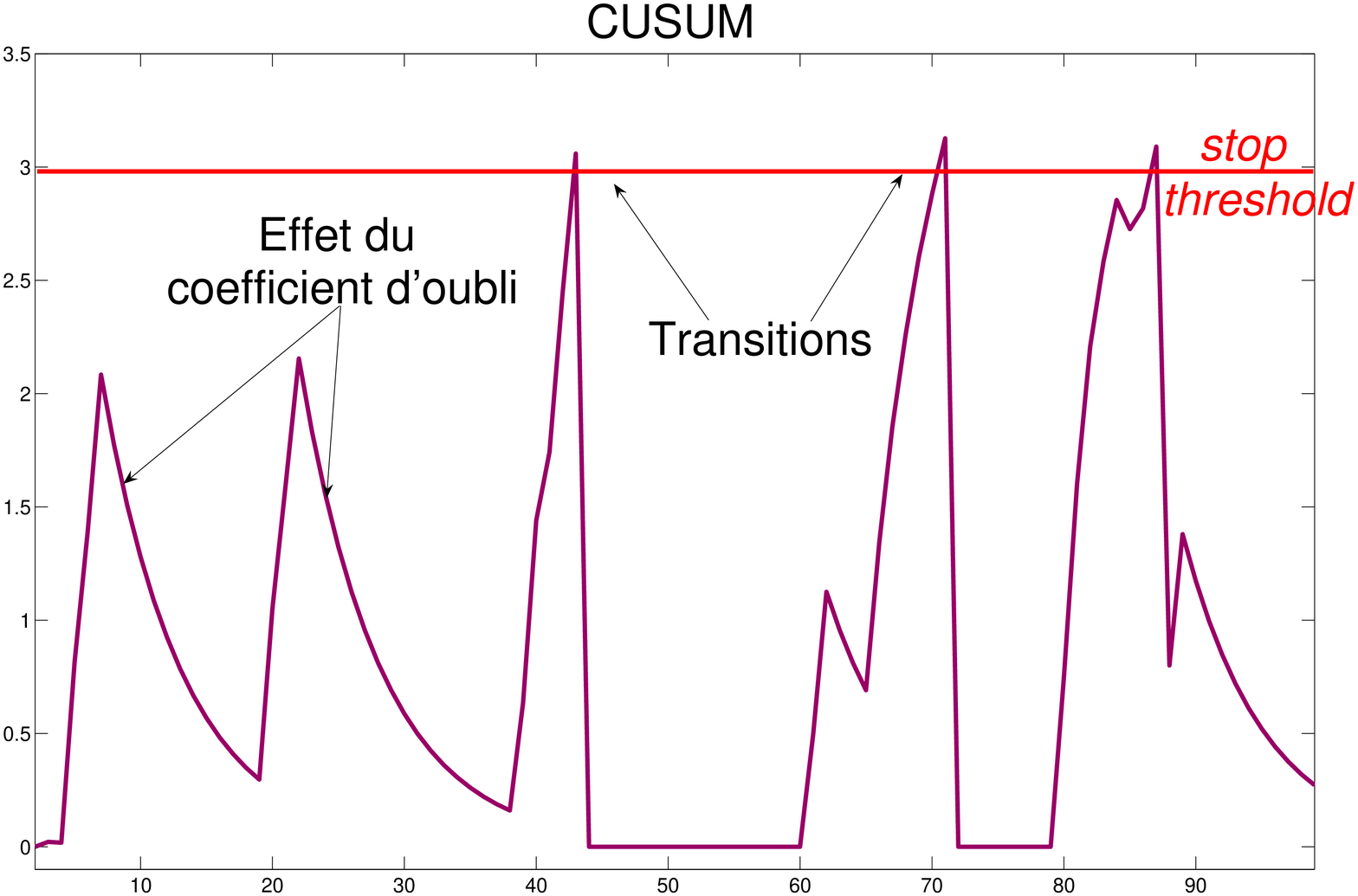}}
\vspace{-0.5cm} \caption{Les croyances avant et après application
du filtre temporel de croyances, et visualisation de la
\textsc{cusum} pour une action \emph{saut} dans un
 saut en hauteur.} \label{fig:exact}
\vspace{-0.5cm}
\end{center}
\end{figure*}

Le tableau~\ref{tab:respvhj} présente les résultats de
reconnaissance des actions pour toutes les séquences avant et
après filtrage par le filtre temporel crédibiliste. Un gain
important peut être observé pour toutes les actions grâce au
filtrage. La figure~\ref{fig:exact} illustre un résultat du filtre
sur une distribution de masse concernant une action \emph{saut}
dans un saut en hauteur. La croyance sur les différentes
propositions est représentée : $\emptyset$ (paramètres discordants
pour $A$), $R_A$ ($A$ est vraie), $F_A$ ($A$ est fausse) et $R_A
\cup F_A$ (doute sur l'état de $A$). L'accumulation du conflit
(\textsc{cusum}) est aussi illustrée soulignant l'effet du
coefficient d'oubli pendant les fausses alarmes.\vspace{-0.2cm}

\begin{table}[ht]
\begin{center}
\caption{Rappel ($\mathcal{R}$) et précision ($\mathcal{P}$) en $\%$ avant et après filtrage pour les
actions \emph{course}, \emph{saut} et \emph{chute} dans des sauts à la perche et des sauts en hauteur.
La colonne de droite donne le gain apporté par le filtre.}
\begin{scriptsize}
\begin{tabular}{|c|c|c|@{~}c@{~}|}
\hline
\textsc{\textbf{Perche}} & \textbf{avant} & \textbf{après} & \textbf{gain} \\
\hline
course    &  $83.8\;\;\;\;71.0$      &      $91.7\;\;\;\;67.9$      & {\footnotesize{(+)}}$7.90\;\;\;\;${\footnotesize{(-)}}$3.1$\\
saut        &  $40.2\;\;\;\;94.7$      &    $78.4\;\;\;\;95.3$      &   $\;${\footnotesize{(+)}}$38.2\;\;\;\;${\footnotesize{(+)}}$0.6$\\
chute     &  $51.5\;\;\;\;87.0$      &      $67.5\;\;\;\;83.8$      &   {\footnotesize{(+)}}$16.0\;\;\;\;${\footnotesize{(-)}}$3.2$\\
\hline
\textbf{moyenne}  &  $\mathbf{63.4 \;\;\;\;77.3}$      &   $\mathbf{82.3\;\;\;\;77.0}$      &
$\;\;${\footnotesize{\textbf{(+)}}} $\mathbf{18.9}\;\;\;$ {\footnotesize{\textbf{(-)}}} $\mathbf{0.3}$\\
\hline
\hline
\textsc{\textbf{Hauteur}} & \textbf{avant} & \textbf{après} & \textbf{gain} \\
\hline
course    &  $89.5\;\;\;\;86.1$      &      $99.8\;\;\;\;84.1$      &   {\footnotesize{(+)}}$10.3\;\;\;\;${\footnotesize{(-)}}$2.0$\\
saut        &  $80.1\;\;\;\;79.7$          &    $84.2\;\;\;\;79.0$      &   {\footnotesize{(+)}}$4.10\;\;\;\;${\footnotesize{(-)}}$0.7$\\
chute     &  $95.5\;\;\;\;91.0$      &      $97.8\;\;\;\;90.1$      &   {\footnotesize{(+)}}$2.30\;\;\;\;${\footnotesize{(-)}}$0.9$\\
\hline
\textbf{moyenne}  &  $\mathbf{88.2\;\;\;\;85.6}$       &    $\mathbf{95.8\;\;\;\;84.0}$     &   $\;${\footnotesize{\textbf{(+)}}}$\mathbf{7.60}\;\;\;\;${\footnotesize{\textbf{(-)}}}$\mathbf{1.6}$\\
\hline
\end{tabular}
\label{tab:respvhj}
\end{scriptsize}
\end{center}
\vspace{-0.3cm}
\end{table}

\section{Conclusion}

Cet article propose un filtre temporel crédibiliste appliqué à la
reconnaissance d'actions humaines dans les vidéos. Ce filtre, basé
sur le Modèle de Croyances Transférables, permet de lisser les
croyances et de détecter les changements d'états de l'action
considérée. La technique consiste à utiliser le conflit entre un
modèle d'évolution des croyances et des mesures. Les mesures ont
été obtenues par fusion de croyances issues de paramètres extrais
des vidéos. L'utilisation d'une somme cumulée (\textsc{cucum}) du
conflit permet d'éviter les changements d'états intempestifs.

Dans cette article, les actions ont été considérées indépendantes.
Nous cherchons à présent à intégrer des relations de causalité
entre les actions telles que rencontrées dans les réseaux
évidentiels~\cite{Xu94} et dans les réseaux de Petri
crédibilistes~\cite{Mrecsqaru99}. L'objectif est la reconnaissance
de séquences d'actions, aussi appelées activités.

\remerciements{Ce travail est en partie soutenu par le réseau
d'excellence SIMILAR. Nous remercions l'université de Crète pour les
échanges de données.}
%

\end{document}